\documentclass[pdftex,twocolumn,epjc3,a4paper]{svjour3}

\RequirePackage[T1]{fontenc}
\smartqed  
\usepackage{amsmath}
\usepackage{amssymb}
\RequirePackage{mathptmx}      
\RequirePackage{flushend}
\RequirePackage{mathrsfs}
\RequirePackage{dsfont}
\usepackage{units}
\usepackage{graphicx}
\usepackage[colorlinks = true,
            linkcolor = blue,
            urlcolor  = blue,
            citecolor = blue,
            anchorcolor = blue]{hyperref}
\usepackage{cite}
\usepackage{xcolor} 
\usepackage{gensymb}
\usepackage{upgreek}
\usepackage[switch]{lineno}
\journalname{Eur. Phys. J. C}

\begin{document}

\author
{%
A.~H.~Abdelhameed\thanksref{addr1} \and
G.~Angloher\thanksref{addr1}\and
P.~Bauer\thanksref{addr1}\and
A.~Bento\thanksref{addr1,addr2}\and
E.~Bertoldo\thanksref{t1,e1,addr1}\and
R.~Breier\thanksref{addr13}\and
C.~Bucci\thanksref{addr3}\and
L.~Canonica\thanksref{addr1}\and
A.~D'Addabbo\thanksref{addr3,addr10}\and
S.~Di~Lorenzo\thanksref{addr3,addr10}\and
A.~Erb\thanksref{addr4,addr5}\and
F.~v.~Feilitzsch\thanksref{addr4}\and
N.~Ferreiro~Iachellini\thanksref{addr1}\and
S.~Fichtinger\thanksref{addr7}\and
D.~Fuchs\thanksref{addr1}\and
A.~Fuss\thanksref{addr7,addr8}\and
V.M.~Ghete\thanksref{addr7,addr8}\and
A.~Garai\thanksref{addr1}\and
P.~Gorla\thanksref{addr3}\and
D.~Hauff\thanksref{addr1}\and
M. Je\v{s}kovsk\'{y}\thanksref{addr13}\and
J.~Jochum\thanksref{addr6}\and
J.~Kaizer\thanksref{addr13}\and
M.~Kaznacheeva\thanksref{addr4}\and
A.~Kinast\thanksref{addr4}\and
H.~Kluck\thanksref{addr7,addr8}\and
H.~Kraus\thanksref{addr9}\and
A.~Langenk\"amper\thanksref{addr4}\and
M.~Mancuso\thanksref{t1,e2,addr1}\and
V.~Mokina\thanksref{addr7}\and
E.~Mondragon\thanksref{addr4}\and
M.~Olmi\thanksref{addr3,addr10}\and
T.~Ortmann\thanksref{addr4}\and
C.~Pagliarone\thanksref{addr3,addr11}\and
V. Palu\v{s}ov\'a\thanksref{addr13}\and
L.~Pattavina\thanksref{addr3,addr4}\and
F.~Petricca\thanksref{addr1}\and
W.~Potzel\thanksref{addr4}\and
P. Povinec\thanksref{addr13}\and
F.~Pr\"obst\thanksref{addr1}\and
F.~Reindl\thanksref{addr7,addr8}\and
J.~Rothe\thanksref{addr1}\and
K.~Sch\"affner\thanksref{addr1}\and
J.~Schieck\thanksref{addr7,addr8}\and
V.~Schipperges\thanksref{addr6}\and
D.~Schmiedmayer\thanksref{addr7,addr8}\and
S.~Sch\"onert\thanksref{addr4}\and
C.~Schwertner\thanksref{addr7,addr8}\and
M.~Stahlberg\thanksref{addr1}\and
L.~Stodolsky\thanksref{addr1}\and
C.~Strandhagen\thanksref{addr6}\and
R.~Strauss\thanksref{addr4}\and
I.~Usherov\thanksref{addr6}\and
F.~Wagner\thanksref{addr7,addr8}\and 
M.~Willers\thanksref{addr4}\and
V.~Zema\thanksref{addr3,addr10,addr15}\and
J.~Zeman\thanksref{addr13}
(The CRESST Collaboration)\\
and\\
M.~Br\"utzam\thanksref{addr20}\and
S.~Ganschow\thanksref{addr20}
}

\institute
{%
Max-Planck-Institut f\"ur Physik, D-80805 M\"unchen, Germany 
\label{addr1} \and
Comenius University, Faculty of Mathematics, Physics and Informatics, SK-84248 Bratislava, Slovakia
\label{addr13}\and
INFN, Laboratori Nazionali del Gran Sasso, I-67100 Assergi, Italy 
\label{addr3} \and
Physik-Department and Excellence Cluster Universe, Technische Universit\"at M\"unchen, D-85748 Garching, Germany 
\label{addr4} \and
Institut f\"ur Hochenergiephysik der \"Osterreichischen Akademie der Wissenschaften, A-1050 Wien, Austria 
\label{addr7} \and
Atominstitut, Vienna University of Technology, A-1020 Wien, Austria 
\label{addr8} \and
Eberhard-Karls-Universit\"at T\"ubingen, D-72076 T\"ubingen, Germany 
\label{addr6} \and
Department of Physics, University of Oxford, Oxford OX1 3RH, United Kingdom 
\label{addr9} \and
also at: Departamento de Fisica, Universidade de Coimbra, P3004 516 Coimbra, Portugal 
\label{addr2} \and
also at: GSSI-Gran Sasso Science Institute, 67100, L'Aquila, Italy 
\label{addr10} \and
also at: Walther-Mei{\ss}ner-Institut f\"ur Tieftemperaturforschung, D-85748 Garching, Germany 
\label{addr5} \and
also at: Dipartimento di Ingegneria Civile e Meccanica, Universit\`{a} degli Studi di Cassino e del Lazio Meridionale, I-03043 Cassino, Italy 
\label{addr11} \and
also at: Chalmers University of Technology, Department of Physics, SE-412 96 G\"oteborg, Sweden 
\label{addr15} \and
Leibniz-Institut f\"ur Kristallz\"uchtung, D-12489 Berlin, Germany
\label{addr20} 
}

\thankstext[$\star$]{t1}{Corresponding authors}
\thankstext{e1}{bertoldo@mpp.mpg.de}
\thankstext{e2}{mancuso@mpp.mpg.de}

\title{Cryogenic characterization of a LiAlO$_{2}$ crystal and new results on spin-dependent dark matter interactions with ordinary matter}

\date{Received: date / Accepted: date}

\maketitle

\begin{abstract}

In this work, a first cryogenic characterization of a scintillating LiAlO$_{2}$ single crystal is presented. The results achieved show that this material holds great potential as a target for direct dark matter search experiments. Three different detector modules obtained from one crystal grown at the Leibniz-Institut f\"ur Kristallz\"uchtung (IKZ) have been tested to study different properties at cryogenic temperatures. Firstly, two 2.8~g twin crystals were used to build different detector modules which were operated in an above-ground laboratory at the Max Planck Institute for Physics (MPP) in Munich, Germany.
The first detector module was used to study the scintillation properties of LiAlO$_{2}$ at cryogenic temperatures.
The second achieved an energy thre\-shold of (213.02$\pm$1.48) eV which allows setting a competitive limit on the spin-dependent dark matter particle-proton scattering cross section for dark matter particle masses between 350 MeV/c$^{2}$ and 1.50 GeV/c$^{2}$. Secondly, a detector module with a 373 g LiAlO$_{2}$ crystal as the main absorber was tested in an underground facility at the Laboratori Nazio\-nali del Gran Sasso (LNGS): from this measurement it was possible to determine the radiopurity of the crystal and study the feasibility of using this material as a neutron flux monitor for low-background experiments.


\end{abstract}

Compiled on \today\

\section{Introduction}

In the past few decades, great effort has been devoted to the investigation of dark matter~\cite{Arcadi2017}. One path which could lead to the identification of this elusive particle(s) is that of direct detection experiments. The goal of most experiments in this class is to detect interactions of a dark matter particle with nuclei of a target material~\cite{Undagoitia2015}.
The CRESST (Cryogenic Rare Event Search with Superconducting Thermometers) experiment, like most other direct searches, has primarily focused on probing spin-independent dark matter-nucleus interactions~\cite{Angloher2015}. CRESST~\cite{CRESST2019} is currently operating CaWO$_4$ and Al$_2$O$_3$ crystals at cryogenic temperatures in the LNGS underground laboratory located below the Gran Sasso massif in Italy. One advantage of this experiment is that the technology is not necessarily tied to the target employed; it is relatively easy to change the target crystal and thereby take advantage of the properties of different target nuclei. \\
In 2019, the CRESST Collaboration published the first results obtained with a li\-thium-based crystal operated in an above-ground laboratory~\cite{CRESSTLi}, showing great potential for dark matter searches using lithium-containing crystals. Lithium is an attractive material because it is the lightest element that can be tested with the CRESST technology, which consists of a scintillating crystal equipped with a tungsten based Transition Edge Sensor (TES) operated at cryogenic temperatures. Since CRESST is heavily oriented towards the search for dark matter particles with sub-GeV mass, the adoption of crystals containing light elements can boost this exploration due to the favorable kinematics.
Furthermore, lithium is one of the best elements to investigate spin-depen\-dent interactions, being
mainly constituted of $^7$Li (92.41 \% natural abundance~\cite{chimica}), which  has $J_N = 3/2$ and $\langle S_{p}\rangle = 0.497$~\cite{pacheco1989nuclear}. We do not investigate spin-depen\-dent interactions with $^6$Li because of the current lack of $\langle S_{p/n}\rangle$ values in the available literature. \\
Another appealing property of these crystals is the possibility to detect the $^6$Li(n,$\upalpha$)$^3$H reaction:
\begin{linenomath}
\begin{equation}
\label{ncapture}
^6\mathrm{Li + n }\rightarrow \alpha~\mathrm{+ ^3H + 4.78~MeV}.
\end{equation}
\end{linenomath}
In fact, one of the most challenging sources of background for a direct dark matter search experiment are neutrons which, like dark matter particles, induce nuclear recoils. Through the detection of the above reaction, which shows a distinctive signature in a scintillating bolometer~\cite{Barinova2010, Cardani2013}, it is viable to precisely measure the neutron flux inside the experimental setup and, with the support of Monte Carlo simulations, it might be possible to reconstruct the energy spectrum of the neutrons. \\
There are many crystals containing lithium that can be employed, such as Li$_{2}$MoO$_4$~\cite{Barinova2010, Casali2013}, Li$_2$Mg$_2$(MoO$_4$)$_3$~\cite{Martinez2012}, \\Li$_2$WO$_4$~\cite{LiWO}, and LiF~\cite{Miuchi02, Danevich2018}.
Amongst these, a crystal with promising properties is LiAlO$_2$.
First, the CRESST technology for the direct deposition of a TES on the crystal surface can be applied. Second, LiAlO$_2$ is a scintillator at room temperature and shows a light emission with a \unit[340]{nm} peak~\cite{Yanagida2017} at which the CRESST light detectors have a high absorption~\cite{Rothe2018}. Finally, Li\-AlO$_2$ also contains $^{27}$Al (100.0\% natural abundance~\cite{chimica}), another interesting element to study spin-dependent interactions, with $J_N = 5/2$ and $\langle S_{p}\rangle = 0.343$ ~\cite{Engel1995}.
The crystal used to build the detector modules operated in this work was produced at the Leibniz-Institut f\"ur Kristallz\"uchtung (IKZ) in Berlin and Section~\ref{sec:2} is dedicated to a summary of the growth procedure. The following sections respectively detail the experimental setup, the data collected for the cryogenic characterization, the neutron and radiopurity measurements, and the dark matter results.

\section{Crystal growth}
\label{sec:2}

All the detector modules used in this work are based on LiAlO$_2$ targets obtained from one single crystal grown at IKZ. The original crystal had a 5~cm diameter and was produced using the Czochralski technique~\cite{COCKAYNE1981}. The primary challenge for the growth of this kind of material stems from its high melting temperature of 1780$\degree$C, which entails a stro\-ng Li$_{2}$O evaporation. Li$_{2}$O evaporates not only from the melt, but also from the growing crystal: in unfavorable thermal conditions, this evaporation is so strong that an Li-free shell of $\alpha$-Al$_{2}$O$_{3}$, a few millimeters thick, can form around the LiAlO$_2$ crystal. To avoid crystal decomposition which would arise from this effect, the axial thermal gradient in the setup must be kept as steep as possible. However, a steep temperature gradient implies an increased superheating of the melt associated with an intensified Li$_{2}$O evaporation from the melt itself: this shifts the melt composition from the desired one towards an Al$_{2}$O$_{3}$-rich melt. Because of non-identical melt and crystal compositions, the crystallization with Al$_{2}$O$_{3}$-rich melt involves solute segregation. To a certain extent, this results in the degradation of the grown crystals, in the form of a non-uniform macro distribution of the constituting elements and/or micro-inhomoge$\-$neities like se\-cond-phase inclusions, mainly LiAl$_{5}$O$_{8}$, due to reduced interface stability.
There is no perfect set of growth conditions and parameters which can avoid all the effects of Li$_{2}$O evaporation: a practical solution will necessitate a compromise among crystal perfection, crystal size, and cost of the process.\\
The crystal used in this work was grown inside a cylindrical iridium crucible of 100~mm diameter in an argon protective atmosphere. The raw materials used for the crystal production are Li$_{2}$CO$_{3}$ and Al$_{2}$O$_{3}$ compounds with a 4N/5N purity. Special attention was paid to the preparation of the raw material in order to prevent Li$_{2}$O losses before the crystal growth: these materials were weighed and mixed in a stoichiometric ratio and calcinated at temperatures between 700$\degree$C and 750$\degree$C in platinum crucibles. The temperature and duration for this preparation was deduced from thermo-gravimetric measurements of the starting materials~\cite{BERTRAM2004189}.

During the crystal growth, the axial temperature gradient was increased step-wise by changing the thermal insulation, until the opaque Al$_{2}$O$_{3}$ shell disappeared entirely and a shiny transparent crystal was obtained. This was achieved by applying a pulling rate of 1.5~mm/h when growing along the (100) direction, together with a crystal rotation between 10 and 25~rpm to improve the melt mixing. A more detailed description of the growth procedure, crystal defects, and tuning of the parameters can be found in~\cite{VELICKOV2008}.

\section{Experimental setup at Max Planck Institute}

Two (20x10x5) mm$^3$ crystals with a mass of 2.8~g each were cut from the LiAlO$_2$ single crystal produced at IKZ, and were used to assemble two different detector modules.

\begin{figure}
\centering
\includegraphics[width=.47\textwidth]{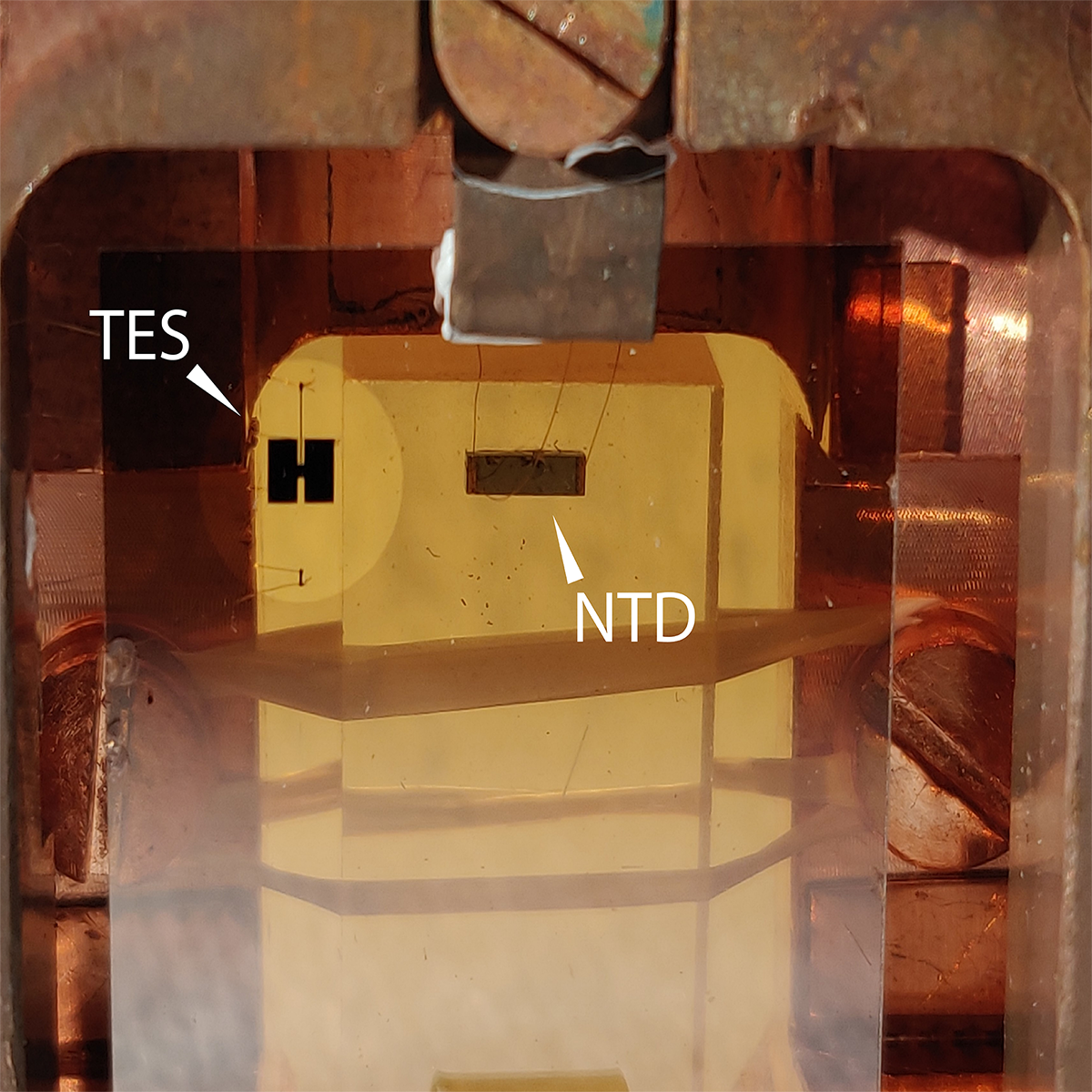}
\caption{Close-up of \textit{module A}. It is possible to see the 2.8~g LiAlO$_2$ crystal instrumented with the NTD sensor through the CRESST-III light detector.}
\label{fig:det_picA}
\end{figure}

\begin{figure}
\centering
\includegraphics[width=.47\textwidth]{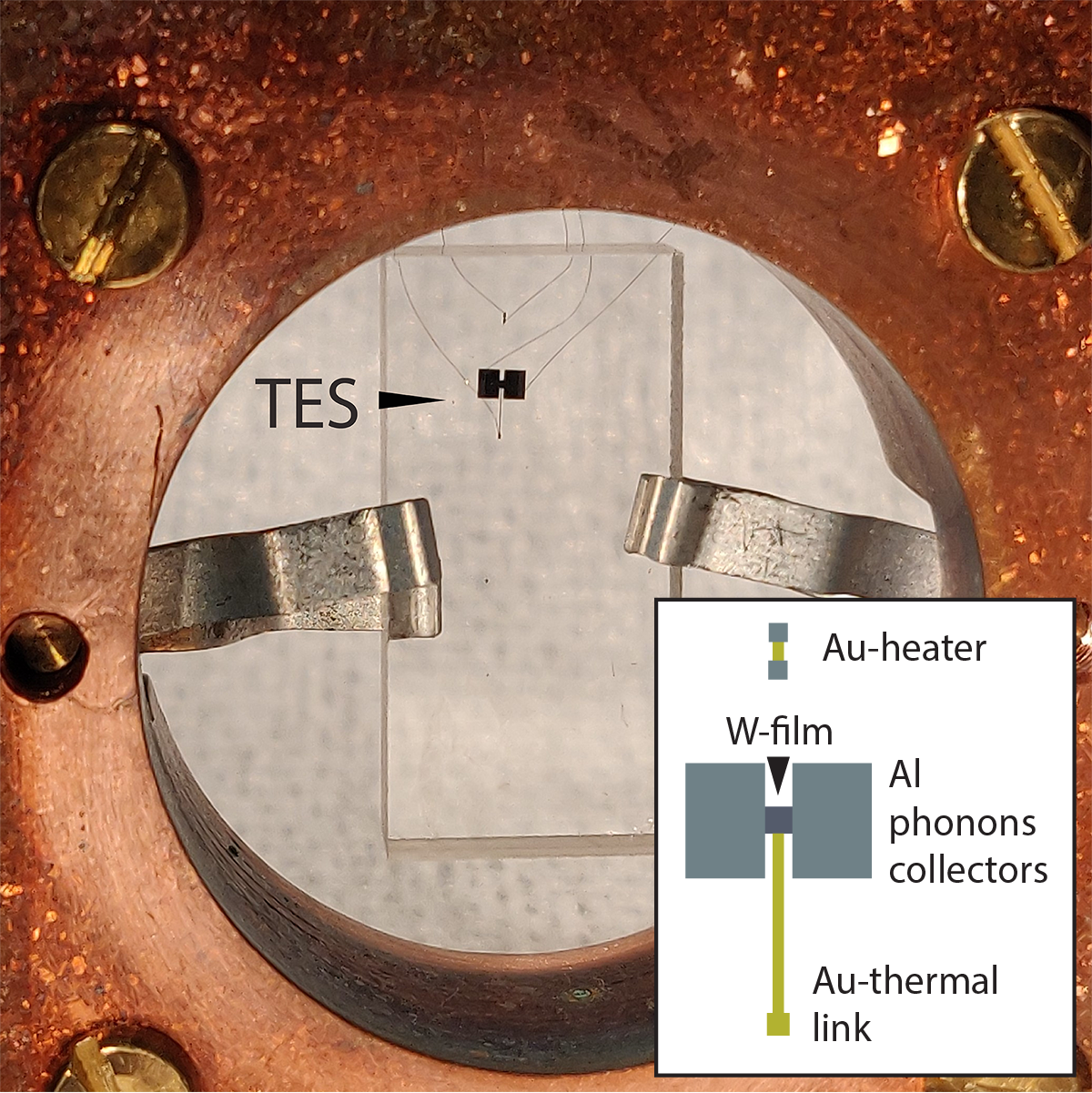}
\caption{Close-up of \textit{module B}, constituted by a 2.8~g LiAlO$_2$ crystal instrumented with a TES directly deposited on the surface. On the bottom right corner a scheme of the TES is shown (see text for details).}
\label{fig:det_picB}
\end{figure}

\begin{figure}
\centering
\includegraphics[width=.47\textwidth]{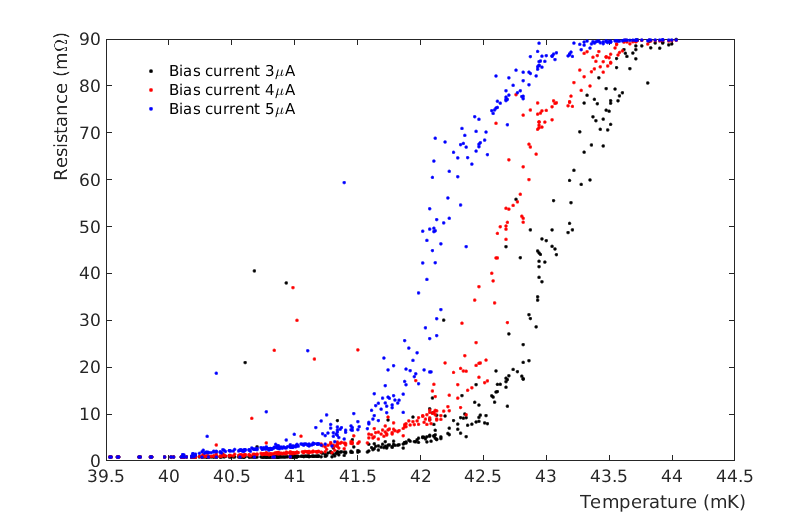}
\caption{Measurement of resistance versus temperature with 3 different bias currents applied to the TES on \textit{module B}. At $\sim$42.5~mK there is a transition between a superconducting response and a normal conducting response (Critical Temperature $T_{C}$) measured with a 4~$\upmu$A bias current. The $T_{C}$ has a slight dependence on the bias current applied to the TES caused by the electrothermal feedback.}
\label{fig:sweep}
\end{figure}

The first crystal was used to assemble \textit{module A}, a detector module (see Figure \ref{fig:det_picA}) designed to characterize LiAlO$_2$ at cryogenic temperatures. In this case, the crystal was instrumented with a Neutron Transmutation Doped (NTD) germanium thermistor~\cite{NTD} glued~\footnote{GP 12 Allzweck-Epoxidkleber, G{\"o}{\ss}l + Pfaff} to one surface. The crystal was held in position inside a copper frame by two strings of PTFE tape. Electrical and thermal connections to the NTD were provided via 25~$\upmu$m diameter gold bond wires.
The temperature variation of the NTD was obtained by measuring the resistance of the thermistor. To do so, a constant bias current was sent through the NTD and the voltage drop of the sensor was measured with a commercial differential voltage amplifier~\footnote{Stanford Research System - SR560 Low-noise voltage preamplifier}. A CRESST-III light detector (LD)~\cite{Rothe2018} was facing the crystal, held in position by two CuNi clamps; this LD was made of a sapphire plate with a 1~$\upmu$m silicon layer epitaxially grown on one face (Silicon-on-Sapphire) with a TES as thermal sensor deposited on the silicon side. The readout of the light detector is obtained with a commercial SQUID system~\footnote{Applied Physics System model 581 DC SQUID}, combined with a CRESST-like detector control system~\cite{ANGLOHER2009}. An $^{55}$Fe X-ray source with an activity of $\sim$\unit[0.05]{Bq} was placed at a distance of $\sim$\unit[0.5]{cm} from the light detector to calibrate its energy response. The TES on the LD had a critical temperature ${T_{C}}^{LD}$=\unit[22]{mK}.\\
The second crystal constituted the main absorber of \textit{mo\-dule B} (Figure \ref{fig:det_picB}), a detector designed to reach a low energy threshold (< 1 keV). The crystal was held in position inside a copper frame by two CuNi clamps. On one face of the crystal, a TES with a design similar to that of the light detector was deposited. The TES is constituted by a thin strip of tungsten with two large aluminum pads partially overlapping the tungsten layer. The aluminum pads serve two different purposes, as phonon collectors and as bond pads. The bond pads are connected via a pair of 25~$\upmu$m aluminum bond wires through which the bias current is injected. The tungsten film is also connected by a long and thin strip of gold to a thicker gold bond pad on which a 25~$\mu$m gold wire is bonded. The gold strip serves as a weak thermal link between the sensor and the heat bath at $\sim$10~mK. On the same surface, but separated from the TES, there is an evaporated heater made of a thin strip of gold with two aluminum pads deposited on top. These pads are bonded with a pair of 25~$\mu$m aluminum bond wires through which a tunable current is injected to maintain the TES at the desired temperature. The heater is also used to inject heater pulses to monitor the detector response over time and for calibration purposes.\\
 This TES had a critical temperature ${T_{C}}^{B}~\simeq$~42.5~mK (Figure \ref{fig:sweep}) when operated with a 4~$\upmu$A bias current. ${T_{C}}^{B}$ is rather high compared to usual transition temperatures of CRESST TESs ($\sim$15~mK); this can negatively affect the performance of the calorimeter, resulting in a higher energy threshold.

The two modules have been operated together inside a Leiden Cryogenics dilution refrigerator at the Max Planck Institute for Physics in Munich, Germany. The dilution refrigerator is located in an above-ground laboratory without shielding against environmental and cosmic radiation. The modules have been mechanically and thermally connected to the coldest point of the dilution refrigerator ($\sim$\unit[10]{mK}).

\section{First cryogenic characterization of LiAlO$_2$}
\label{sec:3}
\begin{figure}
\centering
\includegraphics[width=.47\textwidth]{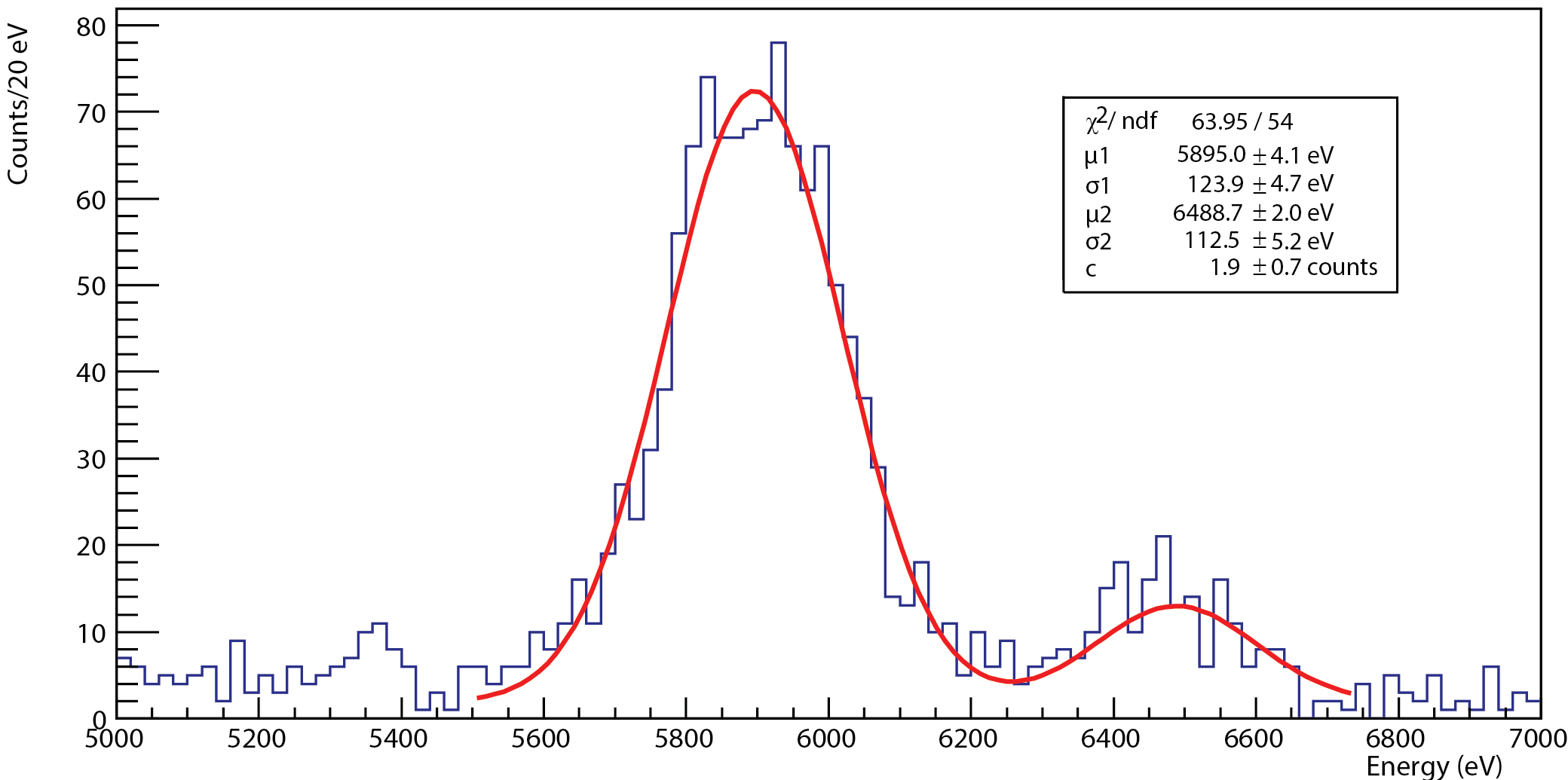}
\caption{Energy spectrum of events registered by the CRESST-III light detector in the energy region of X-ray emission by the $^{55}$Fe source. Two peaks are visible: one at 5.895~keV, resulting from the sum of K$_{\alpha1}$ and K$_{\alpha2}$ lines, and one at 6.490~keV, resulting from the sum of K$_{\beta2}$ and K$_{\beta3}$ lines. The fit function (red solid line) consists of the sum of two Gaussian functions ($\upmu_1$ and $\upmu_2$ are the expected values, $\upsigma_1$ and $\upsigma_2$ the standard deviations) plus a constant factor c to account for the flat background. In principle, $\upsigma_2$ should not be lower than  $\upsigma_1$, but we attribute this anomaly to the presence of an energy loss in the left shoulder of the 5.895~keV peak. The 5.895~keV peak is used to obtain the energy calibration of the light detector.  }
\label{fig:fitluce}
\end{figure}

\begin{figure}
\centering
\includegraphics[width=.47\textwidth]{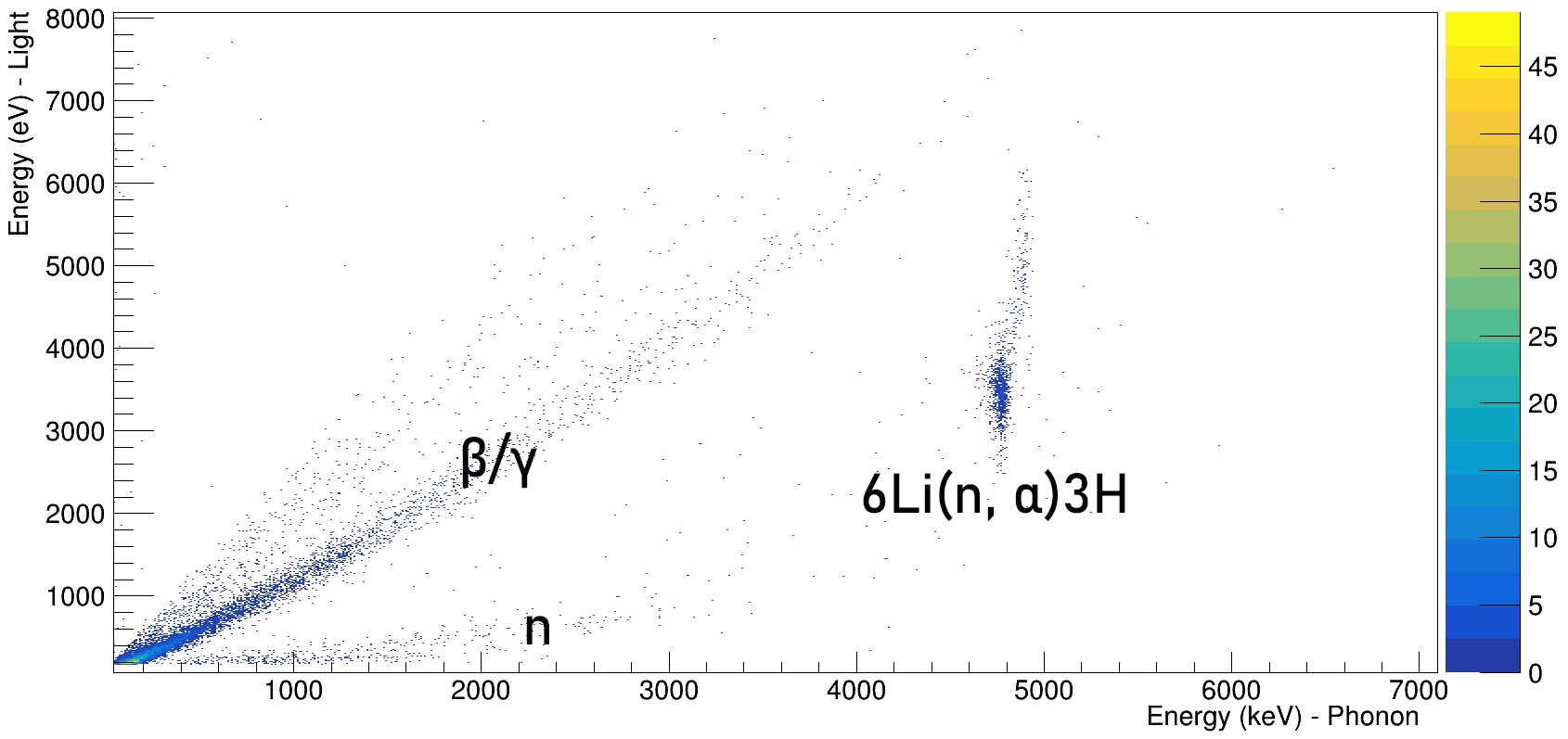}
\caption{Energy measured by the CRESST-III light detector versus energy measured by the NTD for each event registered by \textit{module A} in the presence of an AmBe neutron source during 9.44 hours of data collection. Two bands starting from zero appear: the one with the higher light emission is constituted by $\beta$/$\gamma$ events interacting inside the LiAlO$_2$ crystal, while the one with lower light emission is caused by the scattering of neutrons within the crystal. At 4780 keV a different family of events appears, due to the neutron capture of $^6$Li. In the vicinity of the neutron capture there is an additional family of events, with a slightly higher energy. This family currently is of unknown origin and the modeling of the anomalously high light yield is particularly challenging.}
\label{fig:scatter}
\end{figure}

Since there is no literature available on the cryogenic performance of LiAlO$_2$, the starting point was to study its basic properties. This was done using \textit{module A}, which allowed an initial overview on scintillation, light yield (LY), and Quenching Factors (QFs)~\footnote{The Quenching Factor for the interaction of an arbitrary particle \textit{x} is defined as: $QF_x = LY_x/LY_{\gamma} $}~\cite{Tretyak2009,QF,QF2} for different particle interactions inside the crystal. \\
The energy calibration of the light detector was performed using the peaks originating from the $^{55}$Fe source (Figure~\ref{fig:fitluce}).
After calibration, the baseline resolution of the light detector is $\sigma^{LD}_{baseline}=(26.64\pm1.20)$~eV, while the resolution at 5.895 keV is $\sigma_{Fe}$=(123.9$\pm$4.1)~eV. \\
During the operation of \textit{module A}, an AmBe neutron source was installed at a distance of $\sim$50~cm from the center of the dilution refrigerator. For the energy calibration of the NTD the neutron capture peak appearing at 4780 keV (Equation~\ref{ncapture}) was used, where the energy resolution is \\$\sigma_{capture}$=(19.96$\pm$0.72)~keV.\\
In Figure~\ref{fig:scatter}, the energy measured in the light channel is plotted versus the energy measured in the phonon channel for each event registered by the detector during 9.44~hours of effective measuring time in the presence of the AmBe neutron source. Three main different families of events are easily distinguishable. Gamma and beta particles interacting in the LiAlO$_2$ form one band starting from zero energy and with a light yield of (1.180$\pm$0.103)~keV/MeV, where the LY is defined as:

\begin{linenomath}
\begin{equation}
\label{lightyield}
\mathrm{LY = \frac{Energy_{LD}}{Energy_{NTD}}}
\end{equation}
\end{linenomath}
Neutrons scattering within the crystal exhibit a band star\-ting from zero energy as well, but with a much reduced light yield (0.284$\pm$0.056)~keV/MeV, resulting in a Quenching Factor for neutrons equal to 0.241. At high energies and with a light yield of $\sim$0.75~keV/MeV (or (3.438$\pm$0.227)~keV at 4.78~MeV), in between the $\beta$/$\gamma$ band and the neutron band, the neutron capture by $^6$Li appears. Assuming a linear light emission up to this energy, the QF for the neutron capture is 0.599. \\ The separation between the $\beta$/$\gamma$ band and the neutron band starts to become evident at $\sim$170~keV; thus, in the energy region of interest for dark matter search ($\sim$0-10 keV) it will be unlikely to achieve an effective particle discrimination based on the light yield, even with a substantial improvement of the light collection in comparison to this measurement.
In the vicinity of the neutron capture a small family of events appears, with a slightly higher energy. There is currently no clear interpretation and further investigations will be carried out to understand its origin, which might be tied to a resonance in the cross section for the $^6$Li(n,$\upalpha$)$^3$H reaction. This family of events has not been observed in the LNGS measurement (see Section~\ref{sec:7}), but the neutron source employed in that case had an extremely reduced activity with respect to the above-ground measurement.

\section{Dark matter results}

As explained in Section~\ref{sec:3}, \textit{module B} was designed to study spin-dependent interactions of low-mass dark matter particles with nuclei of a LiAlO$_2$ crystal in a cryogenic measurement.
A low threshold is a key parameter to reach this goal, due to the steep increase of expected dark matter recoils at lower energies. For this reason the
TES was directly evaporated onto the LiAlO$_2$ surface, applying, for the first time, the CRESST technology on a crystal containing lithium .\\
A total of 22.2 hours of data without any source ("background data") were collected for \textit{module B} using a continuous DAQ with a sampling rate of 25~kHz. The events were triggered with a dedicated software based on the optimum filter~\cite{Gatti1986}.

\begin{figure}
\centering
\includegraphics[width=.47\textwidth]{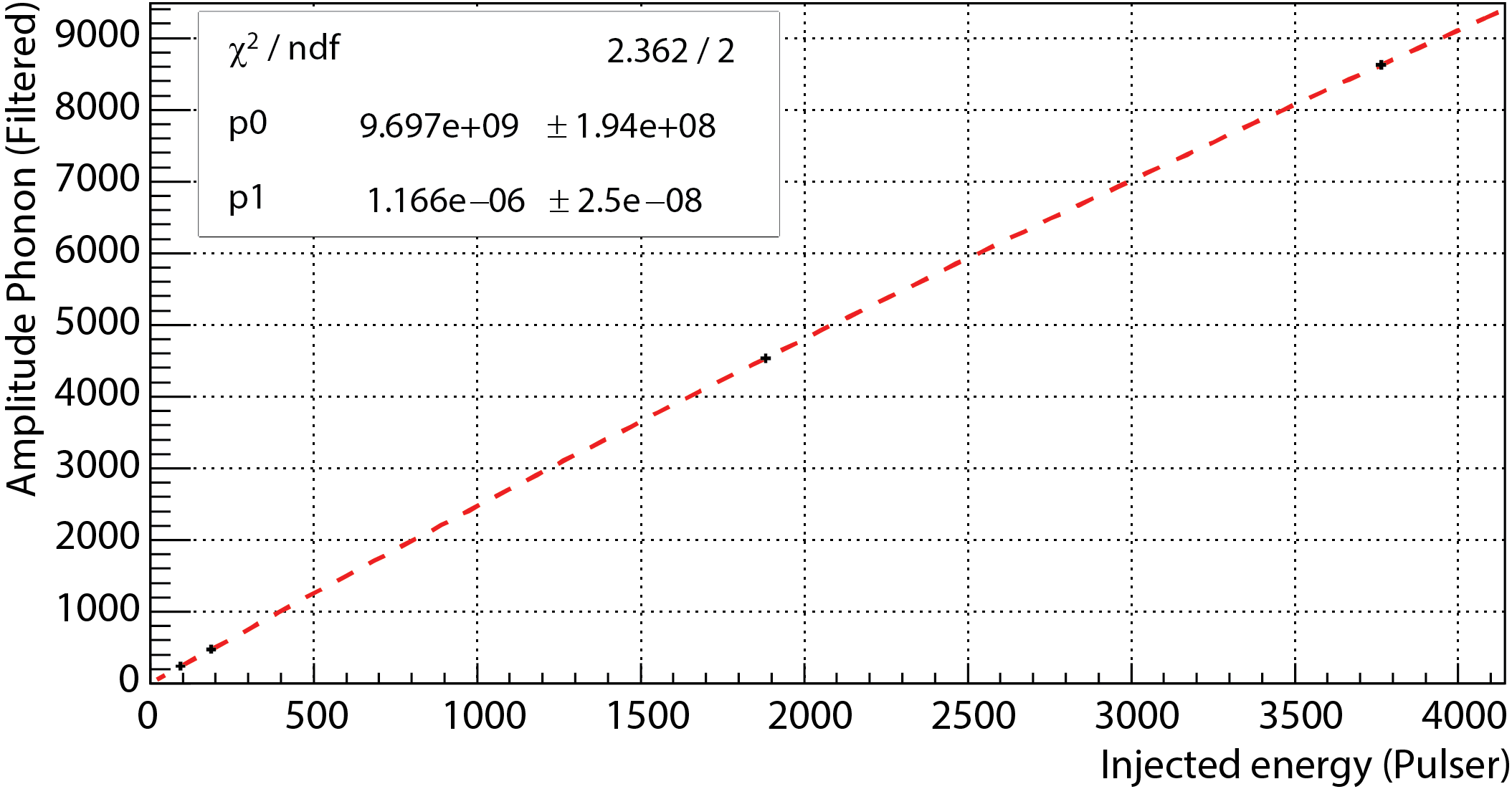}
\caption{Injected amplitude of 4 different heater pulses versus amplitude registered by the TES (black crosses) expressed in arbitrary units. Each amplitude registered by the TES with relative error are obtained from the peaks appearing in the raw spectrum via a Gaussian fit. The four points are fit with Equation~\ref{poly}, which is used for the energy calibration of the detector.}
\label{fig:encal}
\end{figure}

The energy calibration is implemented using the\\5.895~keV peak from the $^{55}$Fe source, similar to the one used for the LD of \textit{module A}. During the run, heater pulses of four different known amplitudes were injected to interpolate the energy calibration in the whole energy region of interest. The peaks corresponding to the heater pulses are identified in the dataset: each peak is fit with a Gaussian function which returns the mean and the error of the mean. Afterwards, the amplitude of heater pulses ($\mathrm{A_{injected}}$) versus the amplitude measured by the TES ($\mathrm{A_{phonon}}$) is plotted and the following function is fit to the data points:
\begin{linenomath}
\begin{equation}
\label{poly}
\mathrm{A_{phonon}}=\mathrm{p0 \cdot \frac{p1 \cdot A_{injected} \cdot I}{R_L + p1 \cdot A_{injected}}}
\end{equation}
\end{linenomath}
where p0 is the gain of the SQUID, I is the bias current of the TES, R$\mathrm{_L}$ is the load resistor, and p1 is a coefficient which translates the temperature change of the TES, induced by the heat pulse, into a variation of the TES resistance. Equation~\ref{poly} is derived from the circuit scheme used to read out the TES~\cite{ANGLOHER2009}. For this measurement I=9~$\upmu$A, while R$\mathrm{_L}$=40~m$\upOmega$; p0 and p1 are the free parameters of the fit. Finally, the mean value registered by the TES  $\mathrm{A_{phonon}}$=(2379.2$\pm$0.7)~a.u. corresponding to the 5.895~keV X-ray is used to convert A$_{phonon}$ to energy via Equation~\ref{poly}. This description assumes that the TES resistivity changes linearly with the temperature in the energy interval considered (0-6~keV). With this method an accurate energy calibration (Figure~\ref{fig:encal}) was obtained taking into account the intrinsic non-linearity of the read-out sche\-me. The baseline resolution is $\sigma^B_{baseline}$=(39.75$\pm$1.23) eV. The corresponding energy threshold for particle interactions with the target is $E^{B}_{T}$=(213.02$\pm$1.48) eV, calculated by using the same method as presented in~\cite{Mancuso:2018zoh}. In this case, however, the total rate of counts in the noise above threshold (noise trigger rate) is set to 10$^3$~counts/(keV$\cdot$kg$\cdot$day), two orders of magnitude lower than the observed event rate in the 1-5~keV range.

\begin{figure}
\centering
\includegraphics[width=.47\textwidth]{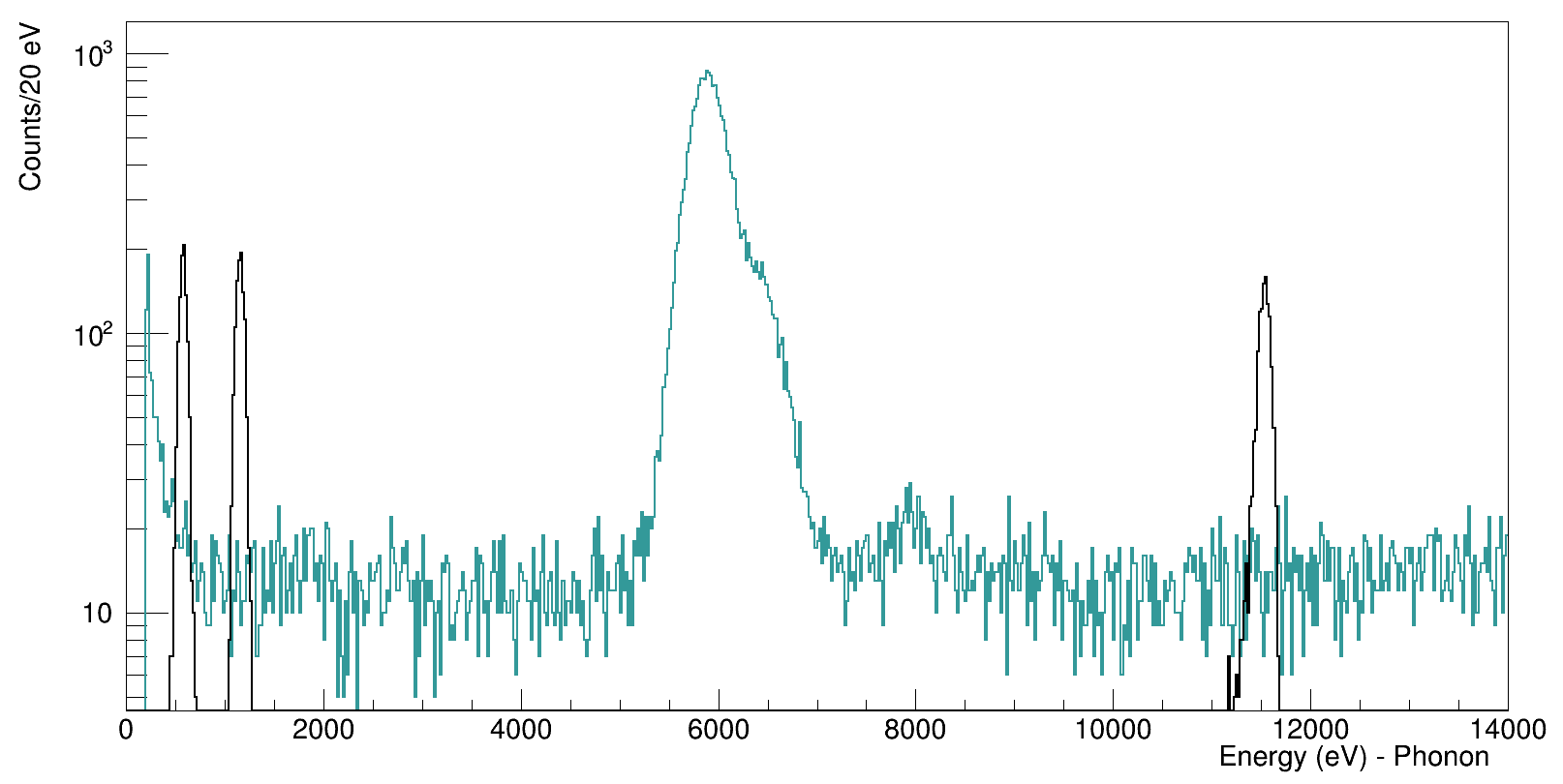}
\caption{Energy spectrum collected during 22.2 hours of background measurement with \textit{module B} without any cut applied to the data set. In black: events induced by injected heater pulses. In light blue: particle events only. At 5.895~keV the peak caused by the X-ray emission of $^{55}$Fe decay appears; the energy resolution at 5.895~keV is equal to $\sigma_{K_{\alpha}}$=(184.0$\pm$1.6)~eV. The resolution for heat pulse events is equal to $\sigma_{HP_2}$=(41.6$\pm$1.0)~eV at 1.159~keV and $\sigma_{HP_3}$=(57.0$\pm$1.6~eV) at 11.537~keV. Below 300~eV there is a rise in the spectrum.}
\label{fig:DMspectrum}
\end{figure}

\begin{figure}
\centering
\includegraphics[width=.47\textwidth]{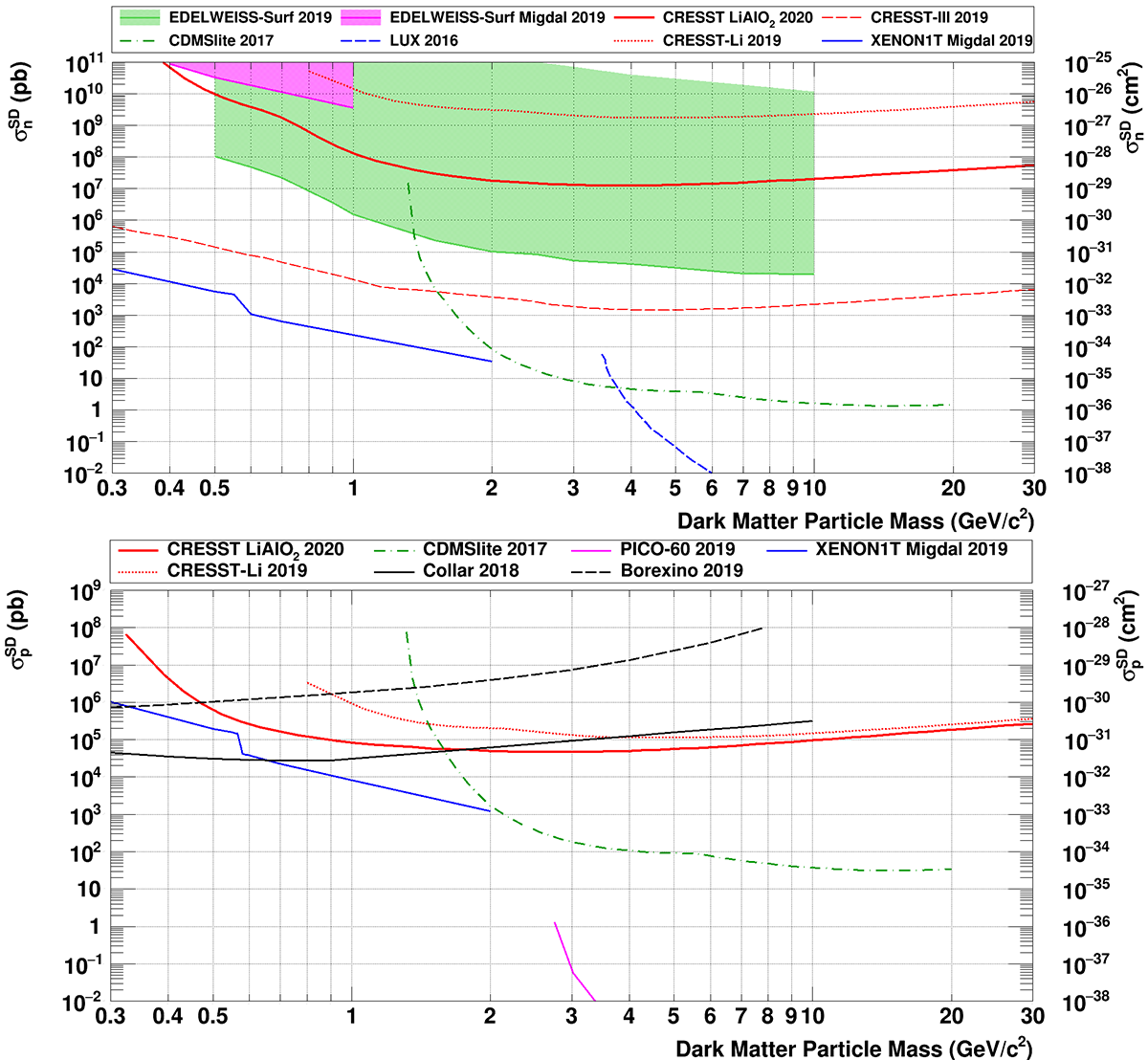}
\caption{\textbf{Top:} Exclusion limits set by various direct detection experiments
for spin-dependent interactions of dark matter particles with neutrons. The result obtained from \textit{module B} data with $^7$Li+$^{27}$Al is shown in solid red. The first result obtained by CRESST using $^7$Li is plotted in dotted red~\cite{CRESSTLi}, while the result obtained with $^{17}$O in CRESST-III is shown in dashed red~\cite{CRESST2019}. For comparison, limits from other experiments are also shown: EDELWEISS~\cite{Armengaud2019} and CDMSlite~\cite{Agnese17} using $^{73}$Ge, LUX~\cite{Akerib2017} and XENON1T (Migdal effect)~\cite{Aprilemigdal} using $^{129}$Xe+$^{131}$Xe.
\textbf{Bottom:} The same, but for spin-dependent interactions of dark matter particles with protons. The result obtained from \textit{module B} data with $^7$Li+$^{27}$Al is shown in solid red. The first result obtained by CRESST using $^7$Li is plotted in dotted red~\cite{CRESSTLi}. Additionally, limits from other experiments are also shown: CDMSlite with $^{73}$Ge~\cite{Agnese17}; PICO with $^{19}$F~\cite{Amole:2019}; XENON1T (Migdal effect) with $^{129}$Xe+$^{131}$Xe~\cite{Aprilemigdal}; Collar with $^{1}$H~\cite{Collar2018}. Finally, a constraint from Borexino data derived in~\cite{Bringmann2018} is shown in dotted black.}
\label{fig:DMLimits}
\end{figure}

Figure~\ref{fig:DMspectrum} shows the calibrated energy spectrum of the 22.2 hours background measurement. The X-ray peaks from $^{55}$Fe decay clearly emerge. A moderate rise of events below 300~eV is also evident. The energy resolution at 5.895~keV, calculated in the same way as for the LD of \textit{Module A}, is $\sigma_{K_{\alpha}}$=(184.0$\pm$1.6)~eV, significantly worse than the resolution for heat pulse events which is equal to \\$\sigma_{HP_2}$=(41.6$\pm$1.0)~eV at 1.159~keV and $\sigma_{HP_3}$=(57.0$\pm$1.6 eV) at 11.537~keV. This degradation in energy resolution for particle events is being investigated further.
In the flat part of the spectrum, below the X-ray peaks, the background rate is of the order of 2$\cdot$10$^5$ counts/(keV$\cdot$kg$\cdot$day), similar to the one observed in~\cite{Angloher2017}. This high value is expected, since the detector is operated in an above-ground laboratory without any shielding or veto systems. \\
From the measured spectrum, dark matter exclusion limits for spin-dependent interactions are calculated. The energy region of interest ranges from $E^{B}_{T}$ to 4000~eV without applying any cut to the particle events registered by the detector.
All the events with energies above 4000~eV contribute to the dead time, reducing the exposure from 22.2 hours to 17.2~hours, corresponding to a total exposure of 2.01$\cdot10^{-3}$ kg$\cdot$day, with an exposure for $^7$Li of 1.95$\cdot10^{-4}$ kg$\cdot$day and an exposure for $^{27}$Al of 8.22$\cdot10^{-4}$~kg$\cdot$day.
The exclusion limits are calculated using Yellin's optimal interval method~\cite{Yellin02, Yellin08} and are shown in Figure~\ref{fig:DMLimits}. The baseline resolution of the detector $\sigma^B_{baseline}$ and the energy threshold $E^{B}_{T}$ are taken into account to evaluate the minimum value of dark matter mass for which it is possible to draw exclusion limits. These limits are valid for both \textit{proton-only} interactions and for \textit{neutr\-on-only} interactions, as discussed in the theoretical fra\-mework presented in~\cite{CRESSTLi}.
The calculation of the exclusion limits adopts the standard dark matter halo model, which assumes a Maxwellian velocity distribution and a local dark matter density of $\rho_\text{DM} = \unit[0.3]{(GeV/ \- c^{2}) /cm^{3}}$~\cite{Salucci2010}. Furthermore, $v_\text{esc} = \unit[544]{km/s}$ is assumed for the galactic escape velocity~\cite{Smith2006} and $v_\odot = \unit[220]{km/s}$ for the solar orbit velocity~\cite{Kerr1986}. For the \textit{proton-only} exclusion limits $\langle S_{p}\rangle = 0.4970$ for $^{7}$Li and $\langle S_{p}\rangle = 0.3430$ for $^{27}$Al are used, while for the \textit{neutr\-on-only} exclusion limits $\langle S_{n}\rangle = 0.0040$ for $^{7}$Li and $\langle S_{n}\rangle = 0.0296$ for $^{27}$Al~\cite{pacheco1989nuclear, Engel1995} are used.
The data analysis efficiency is computed generating  a  known  flat  energy  spectrum  of  events. These events are created by superimposing the ideal detector response on recorded data and then processed with the same analysis chain used for the real data. The fraction of surviving events over the total simulated events at each energy bin represents the data analysis efficiency. Since the determination of the amplitude and the triggering are done in one step by the optimum filter and no further data selection criteria applied, in this case the data analysis efficiency is equivalent to the trigger efficiency.

\section{Experimental setup at LNGS}

\begin{figure}
\centering
\includegraphics[width=.47\textwidth]{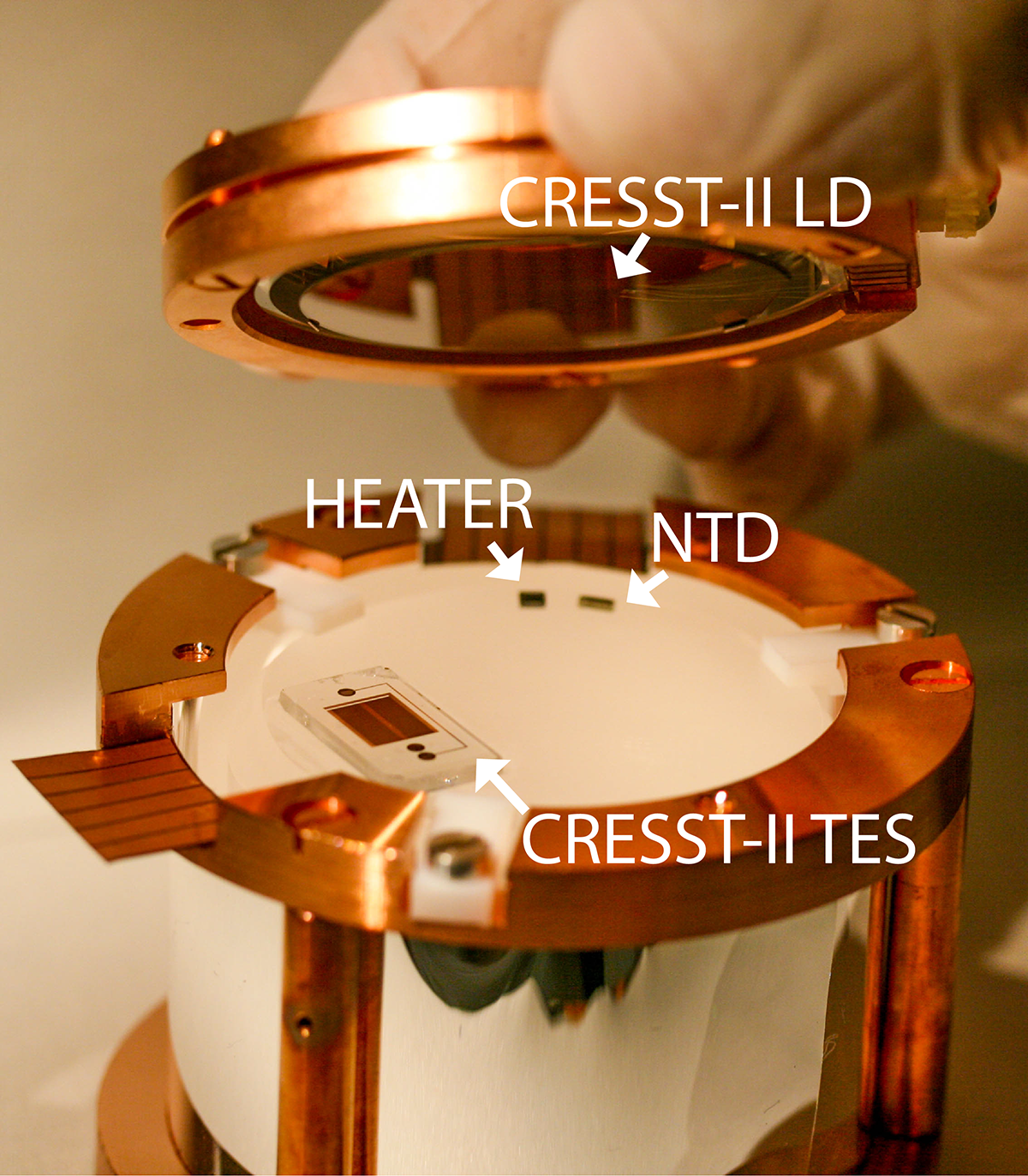}
\caption{Detector \textit{module C} was operated at LNGS. A 373~g crystal is instrumented with two phonon sensors glued on the top surface: an NTD and a CRESST-II TES. On the same surface there is a glued heater which ensures the stability of the detector operation. The crystal is surrounded by reflective foil and a CRESST-II light detector is facing the top surface of LiAlO$_2$.}
\label{fig:biglao}
\end{figure}

After the successful tests at MPP, the bulk of the original LiAlO$_2$ crystal sample was mechanically polished  obtaining a 373~g crystal. Such crystal size is ideal to study the crystal radiopurity and to assess the feasibility of using LiAlO$_2$ crystal as a monitor for the neutron flux in a shielded experimental setup. For this reason, this crystal was used in a new detector module, \textit{module C}, which was installed in the MPP Test-Cryostat facility located in the underground laboratory of Laboratori Nazionali del Gran Sasso (LNGS), under 3600~m water equivalent overburden to shield against cosmic radiation~\cite{Ambrosio:1995cx}.\\
 As visible in Figure~\ref{fig:biglao}, an NTD~\cite{NTD}, a (5$\times$5$\times$1)~mm$^{3}$ Si carrier with a thin gold stripe heater deposited on it, and a CaWO$_4$ carrier crystal on which a CRESST-II TES had been evaporated~\cite{Strauss2014} are both glued~\footnote{\label{note1}GP 12 Allzweck-Epoxidkleber} to the top surface of the LiAlO$_2$ crystal. The NTD and the CRESST-II TES are both being used as phonon sensors. This choice is motivated by the fact that the NTD has a higher dynamic range than the TES, while the TES can generally achieve a lower energy threshold than the NTD. Therefore, with this detector module it is possible to study both the low energy part of the spectrum ($\sim$1~keV) and the high energy part ($\sim$10~MeV). This allows the potential setting of competitive limits for spin-dependent dark matter search and the detailed study of the neutron capture by $^{6}$Li during the same measurement. The crystal was held in position inside a copper holder using three PTFE clamps on the bottom and three on the top. Reflective foil~\footnote{3M's Vikuiti\textsuperscript{TM} Enhanced Specular Reflector} was used to surround the crystal, in order to maximize light collection. A CRESST-II light detector~\cite{Angloher2016} was facing the top surface of the LiAlO$_2$ crystal, completing the detector module. \\
  The MPP Test-Cryostat facility is located in the corridor connecting Hall A and Hall B of LNGS. The model of dilution refrigerator installed in this facility is the same as the one used for the above-ground measurement at MPP. The detector module operated in this dilution refrigerator employs the same kind of wiring, NTD readout, and TES readout as in the previous above-ground measurement.

\section{Neutron and radiopurity measurements at LNGS}
\label{sec:7}
\begin{figure}
\centering
\includegraphics[width=.47\textwidth]{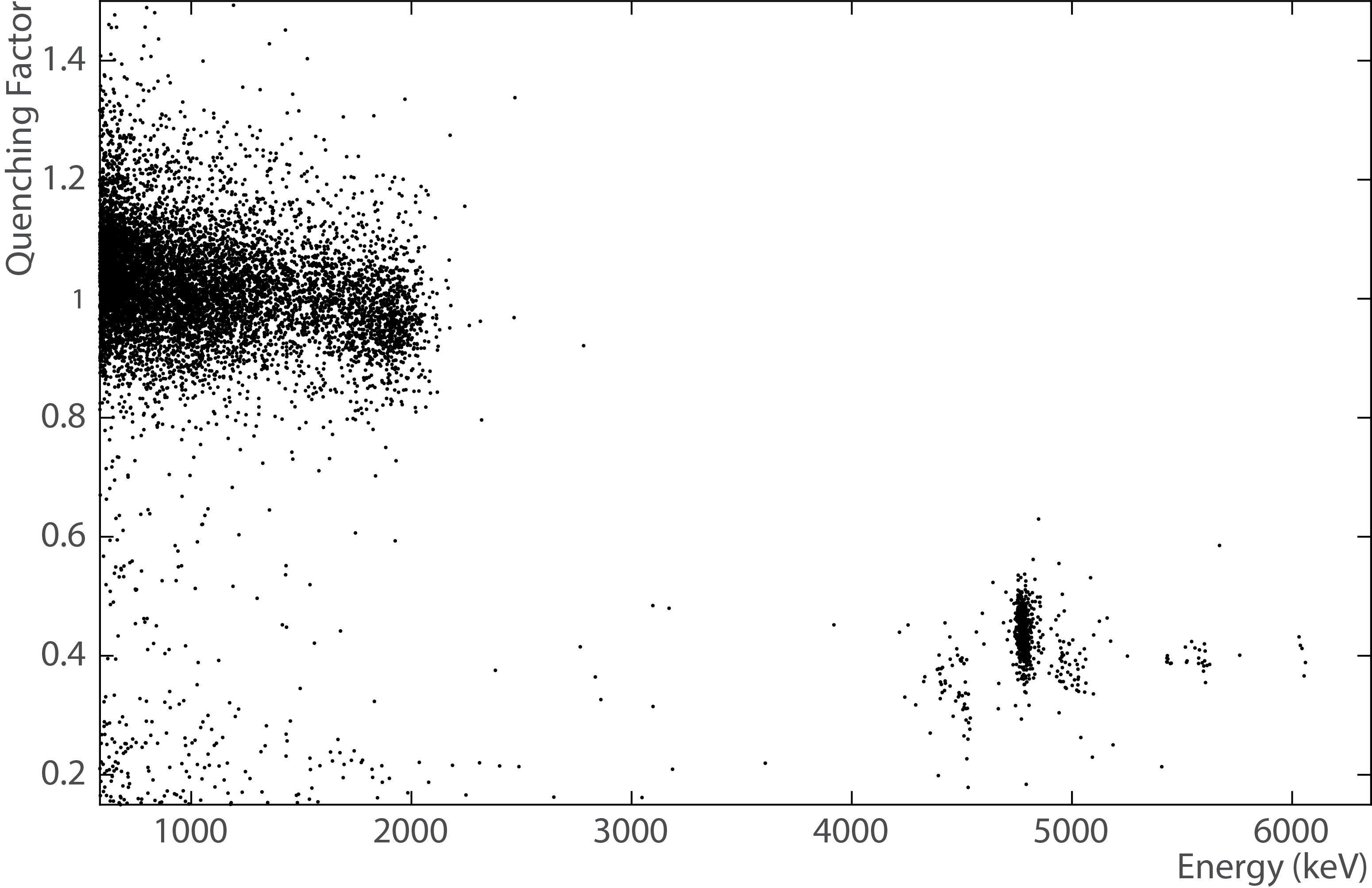}
\caption{QF versus the energy registered by the NTD sensor for 13.1 hours of effective live time in the presence of a weak AmBe source. For energies $\lesssim$2.6~MeV it is possible to see the $\beta$/$\gamma$ band which was used to normalize the QF. At energies $\gtrsim$3~MeV and for lower QF values, multiple families originated by $\alpha$ decays and one prominent line at 4.78~MeV corresponding to the neutron capture of $^6$Li can be seen. }
\label{fig:scatterplot}
\end{figure}

\begin{figure}
\centering
\includegraphics[width=.47\textwidth]{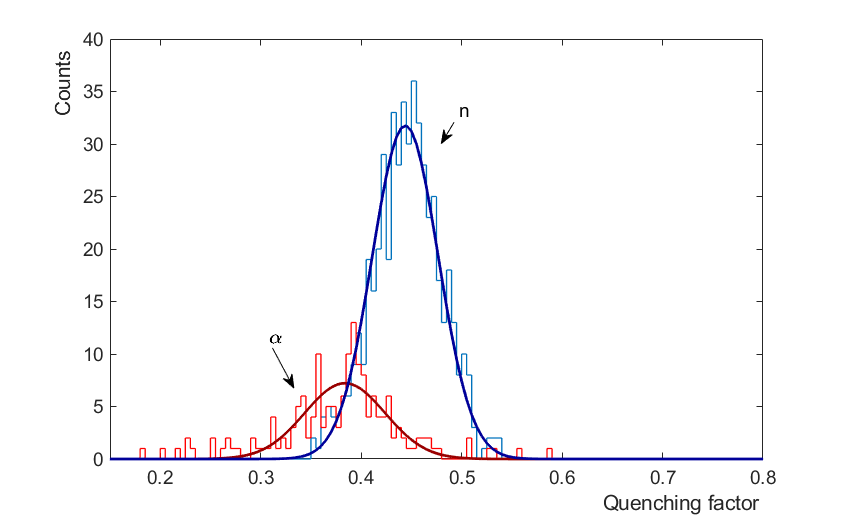}
\caption{Distribution of QF values for events originated by $\alpha$ decays with a mean value of 0.38$\pm$0.04 (red) and distribution of QF for neutron capture events with a mean value of 0.44$\pm$0.03 (blue).}
\label{fig:quenching}
\end{figure}

The detector operation of \textit{module C} at LNGS was divided into two parts: one focused on the efficacy of measuring neutrons, the other on measuring the radioactive impurities in the crystal. For these type of measurements, the data collected with the NTD (that does not saturate in the energy region of interest) and the CRESST-II light detector have been analyzed. The CRESST-II TES was also simultaneously operated as a phonon sensor to study the low-energy part of the spectrum (<1~MeV). \\
At the beginning of the run an AmBe neutron source emitting $\sim$10 neutrons/s was installed at a distance of $\sim$60 cm from the center of the dilution refrigerator and 13.1 hours of data were collected. To ensure the stability of the NTD sensor, heater pulses with seven different amplitudes were injected, two of which were close to the energy region of interest for the neutron capture by $^6$Li (Equation \ref{ncapture}). The detector response was calibrated using these heater pulses and the 4.78~MeV peak corresponding to the neutron capture. After calibration, the energy resolution at 4780 keV is\\ $\sigma_{capture}$=(18.3$\pm$1.02) keV. In Figure~\ref{fig:scatterplot}, the scatter plot of QF versus the energy registered by the NTD for all the events recorded during the neutron measurement is presented.
In this plot, the neutron capture peak shows a higher QF than the events originating from $\alpha$ decays. These two classes of events are used to build two histograms (Figure~\ref{fig:quenching}): neutron capture events are selected from an energy interval of $\pm$3$\sigma_{capture}$ centered around 4780 keV, while all other events above 4~MeV are considered alpha events. It is possible to see that the two distributions are partially overlapping. However, even with a simple cut on the QF value one can exclude the vast majority of unwanted $\alpha$ decay events: if only events with a QF>0.44 (the mean value of the neutron capture distribution) are accepted, 93.3\% of $\alpha$ events are cut while halving the detection efficiency for the neutron capture. The efficiency in discarding $\alpha$ events can then also be considerably increased defining a cut on the energy detected by the NTD phonon sensor: clearly, this cut is more effective the higher the energy resolution of the NTD. \\ In a low-background environment only a few neutron events are expected, while the number of alpha events depends on the radiopurity of the detector.  This means that there would probably be not enough events to build two distributions based on the QF values. However, it is possible to perform a neutron calibration and then, based on the data, define a region where neutron events are expected during the background data campaign. From the total number of events inside this region, it is then feasible to quote a neutron flux value (or upper limit) with the respective uncertainty.

The long term goal for CRESST is to directly detect neutrons inside the experimental setup using a specifically designed detector based on a lithium-containing crystal, thereby providing a relevant input to the background model of the experiment. From this data, using dedicated Monte Carlo simulations, the total neutron flux (or an upper limit) can be assessed while also possibly reconstructing the energy spectrum of the incoming neutrons.
The measurement presented in this work is a first step in this direction.

\begin{figure}
\centering
\includegraphics[width=.47\textwidth]{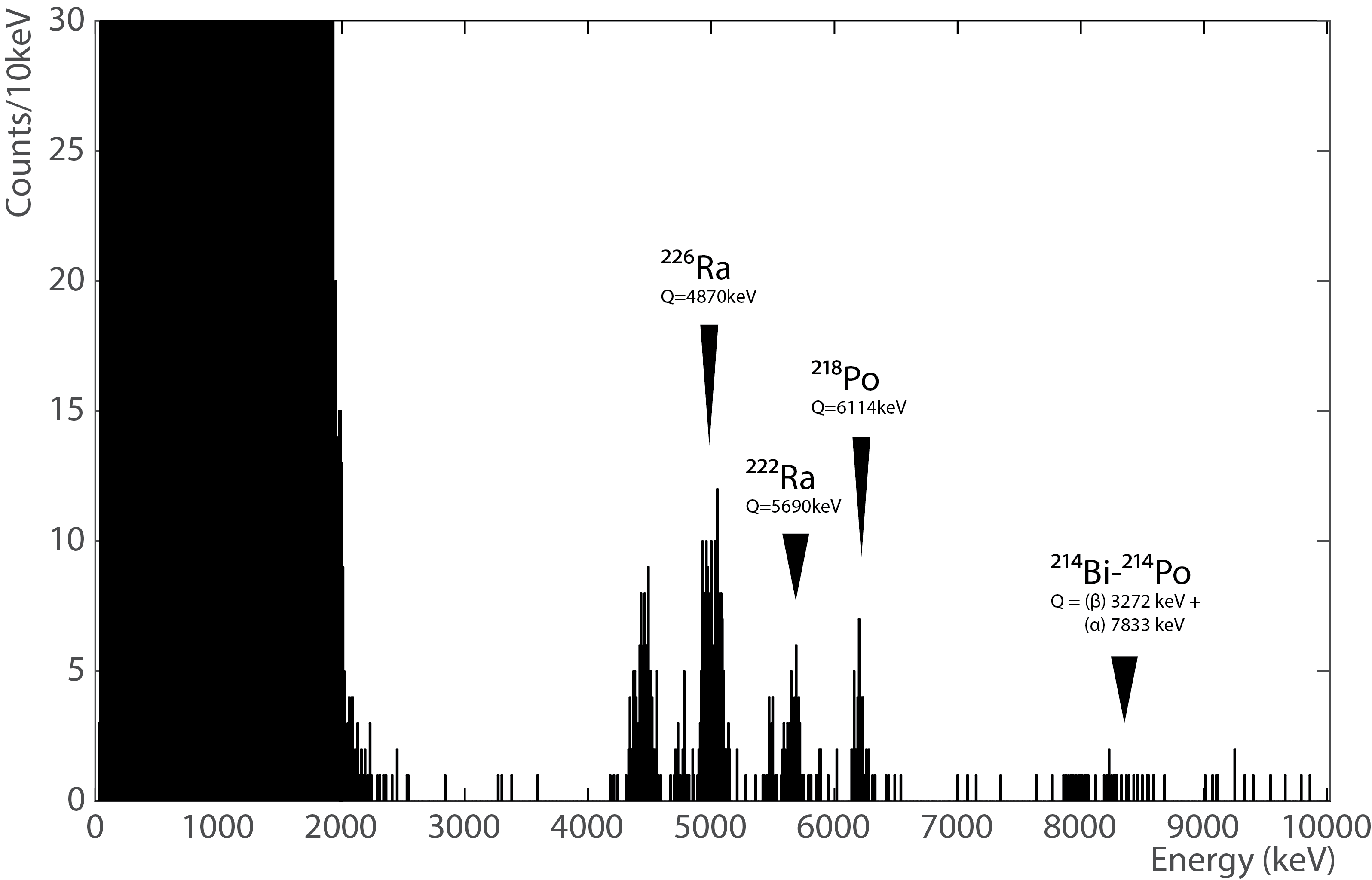}
\caption{Energy spectrum registered by the NTD during a background measurement of 35.6 hours effective time. From this spectrum at least 7 different sources of $\alpha$ decays in the 4-7~MeV region can be distinguished and above 7~MeV additional events, likely due to $^{214}$Bi-$^{214}$Po decays, appear.}
\label{fig:alphas}
\end{figure}

After the neutron measurement, the AmBe source was removed to measure the radiopurity of the crystal. In this case, a 58.4 hours background measurement was carried out. After stability and data quality cuts, the effective measuring time is 35.6 hours. In this measurement it was not possible to use the neutron capture peak to calibrate the NTD response, but the heater pulses that were previously calibrated were used instead. In Figure~\ref{fig:alphas}, the energy spectrum measured by the NTD is shown without cuts. From this spectrum, at least 6 different peaks due to $\alpha$ decays in the 4-7~MeV region can be distinguished. After a careful evaluation, it can be assumed that three radioactive parents are inducing the peaks highlighted: $^{210}$Po, $^{226}$Ra, and natural uranium. The respective calculated activities are listed in Table~\ref{tab:my_label}. In principle, $^{226}$Ra should be part of the $^{238}$U decay chain, but it is not possible to correctly match the respective activities. One straightforward explanation is that $^{226}$Ra and $^{238}$U are not in secular equilibrium; as such the two contaminants are treated as separate parents of their respective decay chain. In fact, in the case of the secular equilibrium the peak centered around 4.86~MeV is expected to be $\sim$3 times more populated than the $^{238}$U peak, due to the summing of $^{226}$Ra, $^{230}$Th, and $^{234}$U activities. However, this peak is only 1.43 times more populated than the $^{238}$U peak and equal, well within 1 sigma, to the sum of $^{222}$Rn (0.962$\pm$0.142~mBq) and $^{238}$U activities. One explanation which can fit well the data is that $^{226}$Ra and its daughters are in secular equilibrium and have the same activities, while we do not observe the daughters of $^{238}$U and $^{235}$U decay chains. The activities ratio of the uranium isotopes are roughly as expected for the presence of natural uranium, only the activity of $^{235}$U is slightly higher than expected, but within 2 sigma. It has to be noted that the $^{235}$U peak is the least populated and so the most affected by statistical uncertainties.
The uranium peaks appear to be broader than the peaks caused by the $^{226}$Ra daughters. This could signal that the uranium might be present both internally and on the surfaces of the crystal, while $^{226}$Ra might prevalently be an internal contamination. This observation, combined with the break of the secular equilibrium between $^{238}$U and $^{226}$Ra, could point at two contaminations at different stages of the crystal production and handling, one related to $^{226}$Ra and one due to natural uranium.\\In addition to the $^{214}$Bi-$^{214}$Po decays, two peaks can be attributed to the daughters of $^{226}$Ra: $^{218}$Po and $^{222}$Rn. Finally, a modest contamination of $^{210}$Po is also observed. \\
\begin{table}[]
    \centering
    \begin{tabular}{c c}
    \hline
         Isotope & Activity (mBq/kg) \\
         \hline
       $^{210}$Po & 0.314$\pm$0.080 \\
       $^{226}$Ra + $^{234}$U & 3.327$\pm$0.257 \\
       $^{235}$U & 0.231$\pm$0.069 \\
       $^{238}$U & 2.260$\pm$0.217 \\
       \hline
    \end{tabular}
    \caption{Activities of the radioactive parents as observed during the background measurement at LNGS. }
    \label{tab:my_label}
\end{table}
The total number of events above 3 MeV is 483: this means an upper bound on the total alpha activity of (10.1$\pm$0.5) mBq/kg for this particular LiAlO$_{2}$ crystal. Considering this value, the radiopurity of this crystal is $\sim$3 times worse than the most radiopure CaWO$_4$ crystal produced within the \\CRESST Collaboration (TUM40)~\cite{Strauss:TUM40}, but in line with standard commercial CaWO$_4$ crystals. The goal for the future is to drastically improve the radiopurity of LiAlO$_{2}$, starting from a careful selection of the raw materials used for the crystal growth, and the material used for cutting and polishing.
Additionally, a 20.8 hours calibration using a $^{241}$Am gamma source installed close to the outer shield of the dilution refrigerator was carried out to test the performance of the CRESST-II TES\cite{bertoldo19}. During the calibration and the background measurement, heater pulses with nine different amplitudes were injected. The 59.54~keV gamma peak from the $^{241}$Am source used for the energy calibration has a resolution of $\sigma_{Am}$=(3.044$\pm$0.074)~keV. Similarly for the TES calibrations presented before, this peak and the peaks corresponding to the injected heater pulses are used to accurately calibrate the detector response at different energies. The sensor has an energy threshold of (2.601$\pm$0.126)~keV, considerably higher than that achieved in the measurement performed above-ground with a smaller LiAlO$_2$ crystal: this is expected due to the large increase in mass as showed by the scaling law described in~\cite{Strauss2017}.

\section{Conclusions}

This work details the results of three different detectors, all of which employ a LiAlO$_2$ target crystal, a material that has never been employed in cryogenic experiments thus far. The cryogenic properties of the material were tested in an above-ground laboratory with a 2.8~g crystal and new limits on spin-dependent dark matter interactions are set with a crystal instrumented with a TES deposited on LiAlO$_2$. A large-size detector with a mass of 373~g was operated in an underground cryogenic facility at LNGS in the presence of a weak neutron source, in order to assess the feasibility to monitor the neutron flux directly inside cryogenic setups.
The results presented in this work demonstrate the high potential of LiAlO$_2$ crystals as cryogenic detectors in the field of low-background applications and contribute to the ongoing search for dark matter.

\begin{acknowledgements}
This work is supported through the DFG by \\SFB1258 and the Origins Cluster, and by the BMBF05A17WO4.

\end{acknowledgements}

\end{document}